\documentclass[manuscript]{aastex}

\newcommand{\fermi}{{\it Fermi}}
\newcommand{\swift}{{\it Swift}}
\newcommand{\integralsc}{{\it INTEGRAL}}
\newcommand{\rxte}{{\it RXTE}}

\shorttitle{When A Standard Candle Flickers}
\shortauthors{Wilson-Hodge et al.}

\begin{document}


\title{When A Standard Candle Flickers}


\author{Colleen A. Wilson-Hodge\altaffilmark{1}, Michael L. Cherry\altaffilmark{2}, 
Gary L. Case\altaffilmark{2}, Wayne H. Baumgartner\altaffilmark{3}, Elif Beklen\altaffilmark{4,5}, 
P. Narayana Bhat\altaffilmark{6}, Michael S. Briggs\altaffilmark{6}, Ascension Camero-Arranz\altaffilmark{7}, 
Vandiver Chaplin\altaffilmark{6}, Valerie Connaughton\altaffilmark{6}, Mark H. Finger\altaffilmark{8},
Neil Gehrels\altaffilmark{9}, Jochen Greiner\altaffilmark{10}, Keith Jahoda\altaffilmark{9}, Peter Jenke\altaffilmark{1,11}, R. Marc Kippen\altaffilmark{12}, Chryssa Kouveliotou\altaffilmark{1}, Hans A. Krimm\altaffilmark{3,13}
Erik Kuulkers\altaffilmark{14}, Niels Lund\altaffilmark{15}, Charles A. Meegan\altaffilmark{8}, Lorenzo Natalucci\altaffilmark{16}, William S. Paciesas\altaffilmark{6},
Robert Preece\altaffilmark{6}, James C. Rodi\altaffilmark{2}, Nikolai Shaposhnikov\altaffilmark{3,17}, Gerald K. Skinner\altaffilmark{3,17}, Doug Swartz\altaffilmark{8}, Andreas von Kienlin\altaffilmark{10}, Roland Diehl\altaffilmark{10}, Xiao-Ling Zhang\altaffilmark{10}}

\email{colleen.wilson@nasa.gov}


\altaffiltext{1}{VP 62 Space Science Office, NASA Marshall Space Flight Center, Huntsville, AL 35812, USA}
\altaffiltext{2}{Department of Physics and Astronomy, Louisiana State University, Baton Rouge, LA, 70803, USA}
\altaffiltext{3}{CRESST/NASA GSFC, Astrophysics Science Division, Greenbelt MD 20771, USA}
\altaffiltext{4}{Physics Department, Middle East Technical University, 06531 Ankara, Turkey}
\altaffiltext{5}{Physics Department, Suleyman Demirel University, 32260 Isparta, Turkey}
\altaffiltext{6}{University of Alabama in Huntsville, Huntsville, AL 35899, USA}
\altaffiltext{7}{National Space Science and Technology Center, Huntsville, AL 35805,  USA}
\altaffiltext{8}{Universities Space Research Association, Huntsville, AL 35805, USA}
\altaffiltext{9}{NASA Goddard Space Flight Center (GSFC), Greenbelt, MD 20771, USA}
\altaffiltext{10}{Max-Planck Institut f\"ur Extraterrestische Physik, 85748, Garching, Germany}
\altaffiltext{11}{NASA Postdoctoral Program Fellow}
\altaffiltext{12}{Los Alamos National Laboratory, Los Alamos, NM 87545}
\altaffiltext{13}{Universities Space Research Association, Columbia, MD 21044, USA}
\altaffiltext{14}{ISOC, ESA, European Space Astronomy Centre (ESAC), PO Box 78, 28691 Villanueva de la Ca\~nada (Madrid), Spain}
\altaffiltext{15}{Danish National Space Center, Technical University of Denmark, Juliane Maries Vej 30, 2100 Copenhagen, Denmark}
\altaffiltext{16}{INAF-IASF Roma, via Fosso del Cavaliere 100, 00133, Roma, Italy}
\altaffiltext{17}{University of Maryland, Astronomy Department, College Park, MD 20742, USA}


\begin{abstract}
The Crab Nebula is the only hard X-ray source in the sky that is both bright enough and steady enough to be easily used 
as a standard candle. As a result, it has been used as a normalization standard by most X-ray/gamma ray telescopes. 
Although small-scale variations in the nebula are well-known, since the start of science operations of the 
\fermi\ Gamma-ray Burst Monitor (GBM) in August 2008,  a $\sim 7$\% (70 mcrab) decline has been observed in the overall Crab Nebula flux in 
the 15 - 50 keV band, measured with the Earth occultation technique. This decline is independently confirmed in the $\sim 15-50$ keV band with three other 
instruments: the \swift\ Burst Alert Telescope (\swift/BAT), the {\it Rossi X-ray Timing Explorer} Proportional Counter Array (\rxte/PCA), 
and the {\it INTErnational Gamma-Ray Astrophysics Laboratory}  Imager on Board \integralsc\ (IBIS). A similar decline is also observed in the $\sim3$ - 15 keV data 
from the \rxte/PCA and in the 50 - 100 keV band with GBM, \swift/BAT, and \integralsc/IBIS. The pulsed flux measured with \rxte/PCA since 1999 is consistent with the pulsar spin-down, indicating that the observed changes are nebular. 
Correlated variations in the Crab Nebula flux on a $\sim3$ year timescale are also seen independently with the PCA, BAT, and IBIS 
from 2005 to 2008, with a flux minimum in April 2007. As of August 2010, the current flux has declined below the 2007 minimum.  
\end{abstract}


\keywords{pulsars:individual: Crab Pulsar, X-rays: individual: Crab Nebula}

\section{Introduction}

X-ray and gamma-ray astronomers frequently consider the Crab supernova remnant to be a steady standard candle suitable as a calibration source \citep[e.g.,][]{1,2,3,50}. \citet{2} presented over 5 years of the Spectrometer on \integralsc\ (SPI, 20 keV - 8 MeV) observations, with fitted flux normalizations at 100 keV consistent with being constant to within the $\sim 3$\% quoted errors. On the basis of data from {\it XMM-Newton}, \integralsc, \swift, {\it Chandra}, \rxte, and 
several earlier missions, \citet{1} have concluded that the Crab flux can be described at least up to 30 keV by the same spectrum proposed by \citet{12} three decades earlier: 
$dN/dE = (9.7 \pm 1.0) E^{-(2.1\pm 0.03)} \rm{photons\ cm}^{-2} {\rm s}^{-1} {\rm keV}^{-1}$, i.e. they describe the Crab as a standard candle. 
 

Driven by the central pulsar's spin-down luminosity, the surrounding remnant consists of a cloud of expanding thermal ejecta and a synchrotron nebula \citep{4} 
with an integrated luminosity $\sim 10^{38}$ erg s$^{-1}$. The pulsar provides a shocked wind that accelerates electrons and positrons to energies $\sim 10^7$ GeV
and a source of kinetic energy driving turbulent motion of a ring of wisps nearly surrounding the synchrotron nebula. High resolution observations reveal wisps and knots moving at velocities up to 0.7~c from 
radio to X-ray energies \citep{14,15,17,13,16}.  A central torus and jet structure 
extending out from the pulsar were observed in X-rays by {\it Chandra} \citep{5}, aligned closely with the pulsar's proper motion \citep{6}. The nebular emission is 
considered to be a combination of synchrotron radiation up to $\sim 100$ MeV and a harder inverse Compton spectrum extending up to TeV energies \citep{7}. 

Observations of the 8 GHz nebular flux in 1985 \citep{19} showed a decrease of $0.167 \pm 0.015$\% yr$^{-1}$, consistent with predictions by \citet{20} for an expanding synchrotron-emitting cloud. 
At optical wavelengths, \citet{21} reported a decrease in the nebula-integrated flux of $0.5 \pm 0.2$\% yr$^{-1}$ from 1987-2002. At X-ray 
energies (2-28 keV), 1996-2002 {\it BeppoSAX} observations described by \citet{22} included a 2\% systematic error to account for the observed fluctuations in time. In the 35 -- 300 keV energy region, \citet{10} reported $\sim 10$ \% variations in the flux observed with the Burst and Transient Source Experiment (BATSE) on the {\it Compton Gamma Ray Observatory (CGRO)}
over periods of days to weeks. 

\citet{23} reported a $\sim 40$\% increase in the unpulsed flux (0.75-30 MeV) measured with the {\it CGRO} Compton Telescope (COMPTEL) between April/May 1991 and August/September 1992. At the same time, \citet{7} reported a $\sim 50$\% decrease in the 75-150 MeV flux and steady emission from 150 MeV to 30 GeV measured with the {\it CGRO} Energetic Gamma Ray Experiment Telescope (EGRET) between 1991 and 1993. They interpret this as steady Compton emission $> 150 $ MeV from long-lived $\sim 5-100$ GeV electrons and $<150$ MeV synchrotron emission from shorter-lived 100 TeV to 1 PeV electrons accelerated by time-variable small-scale shock structures. A change in this electron acceleration mechanism would drop a portion of the electrons from the range responsible for the EGRET emission 
to the COMPTEL range, resulting in the observed fluxes. The \fermi\ Large Area Telescope (LAT) found no variations with time in the 100 MeV -
30 GeV band from 2008 August - 2009 April \citep{Abdo10a}. Recently,  \fermi\ LAT (Abdo et al. 2010b) reported flares from the Crab nebula above 100 MeV in 2009 February and 2010 September. {\it AGILE} simultaneously detected the 2010 flare \citep{57}. Reports by the High Energy Gamma Ray Astronomy experiment \citep[HEGRA,][]{48}, the Major Atmospheric Gamma-ray Imaging Cherenkov telescope \citep[MAGIC,][]{49}, the High Energy Spectroscopic System \citep[H.E.S.S.,][]{55}, and the Very Energetic Radiation Imaging Telescope System \citep[VERITAS,][]{51} provide no evidence for time variability, consistent with expectations for higher energies. 

Although it is extremely difficult to obtain absolute flux measurements with accuracy $\sim 1$ \% with a single instrument, we have analyzed independent data sets from four separate operating missions. 
In Section 2, we present the Crab light curves measured independently by \fermi/GBM, \swift/BAT, \integralsc/IBIS and JEM-X, and \rxte/PCA. In Section 3, 
we summarize the results and discuss their implications.

\section{Observations \& Results}

\subsection{\fermi\ GBM}

\begin{figure}
\plotone{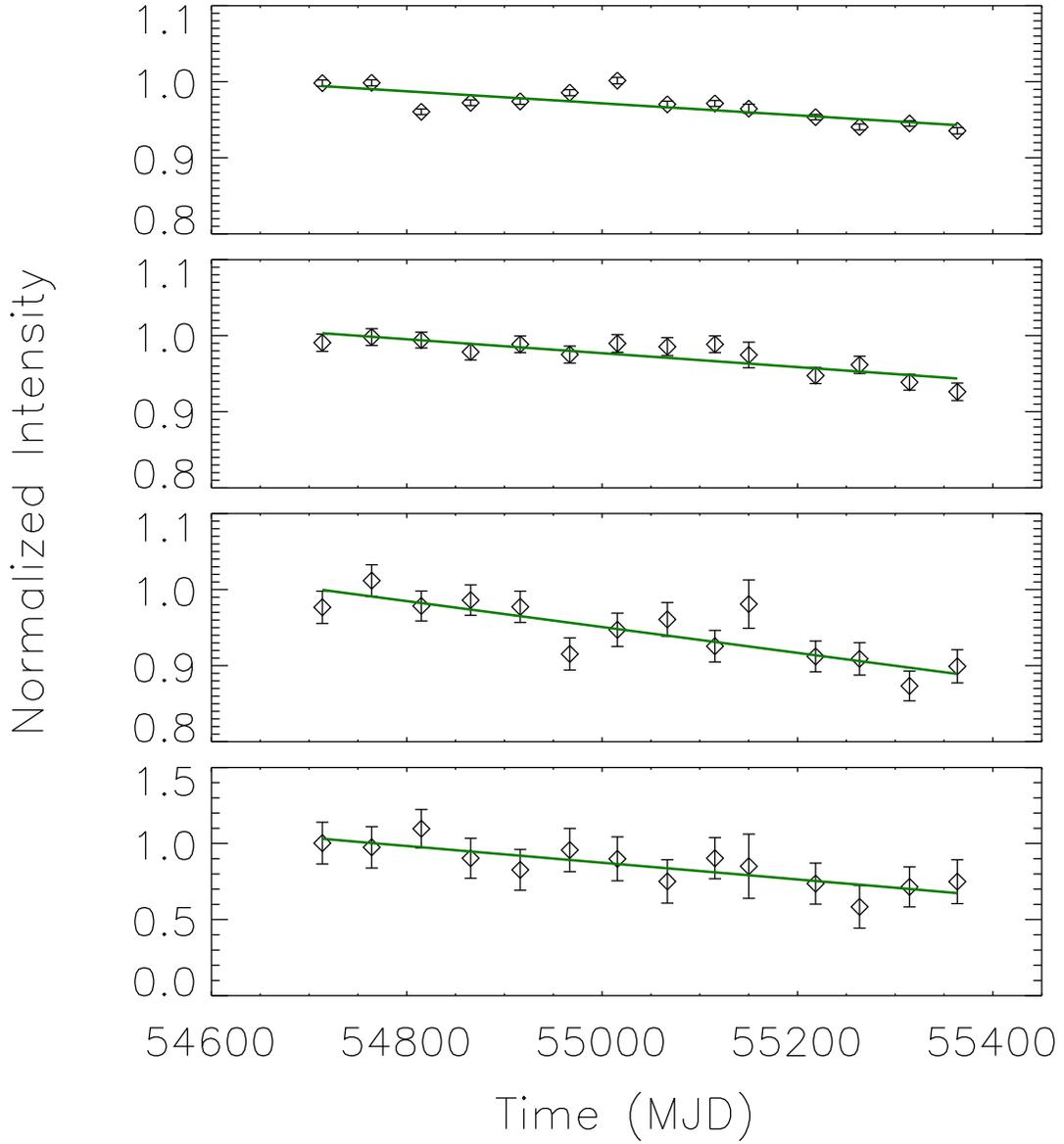}
\caption{From top to bottom: 50-day average GBM Crab measurements for 12-50, 50-100, 100-300, and 300-500 keV. Solid lines are fits used to compute the \% change in rate.\label{gbm}}
\end{figure}

The GBM instrument \citep{24}, sensitive from 8 keV to 40 MeV, provides nearly continuous full-sky coverage via the Earth occultation technique, successfully demonstrated with BATSE \citep{25, 26}. The \citet{25} approach has been adapted for GBM \citep{27, 28}. To date, six persistent and two transient sources have been detected above 100 keV \citep{28}, including the Crab.

The GBM implementation of the Earth occultation technique uses both
CTIME (8 energy channels with 0.256-second resolution) and CSPEC data (128
energy channels with 4.096 s resolution). A detailed detector response model has
been developed based on Geant4 simulations confirmed by extensive ground testing in order to determine the response as a function of orientation \citep{29,56}. In flight, fits to background lines (e.g., 511 keV) over time show a stable gain and energy resolution in all the GBM detectors and electronics, with lines typically 
within 1\% of their expected position. 

The Crab light curve measured in four energy bands with GBM from August 12, 2008 through July 13, 
2010 (MJD 54690-55390) is shown in Figure~\ref{gbm}. With respect to the rate on MJD 54690, the Crab rate appears to have decreased steadily by more than 5\%: The decrease is $5.4\pm0.4$\%, $6.6\pm1.0$\%, $12\pm2$\%, and $39\pm13$\% 
in the 12-50, 50-100, 100-300, 300-500 keV bands, respectively. Inclusion of a linear decline in the 12-50 keV band improves reduced $\chi^2$ to 605.8/130=4.66 from 956.3/131=7.30 for a constant Crab.

\subsection{\rxte\ PCA}

Frequent observations with the \rxte\ PCA were made to monitor the radio-X-ray phase of the Crab pulsed emission \citep{42} and for calibration purposes \citep{43, 44}. In the PCA, the Crab is bright ($\sim$ 2500 counts s$^{-1}$ detector$^{-1}$). Unrejected background from all 
sources amounts to about 1 mCrab. The PCA is a relatively simple instrument, with commanded changes in 
operating conditions limited to the high voltage. Data since the last high voltage change in 1999 for PCU 2,3, and 4 are used in this paper.

\begin{figure}
\plottwo{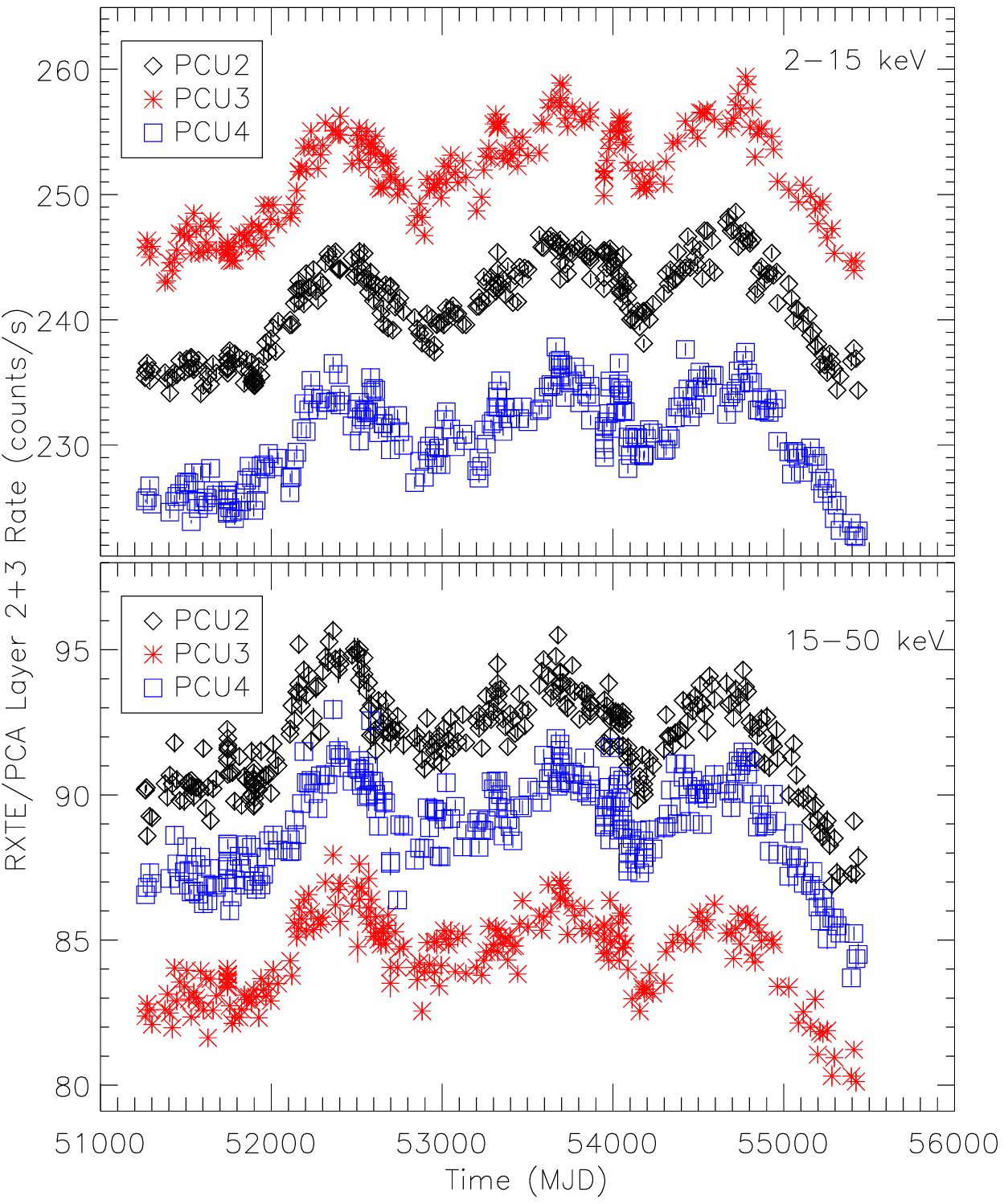}{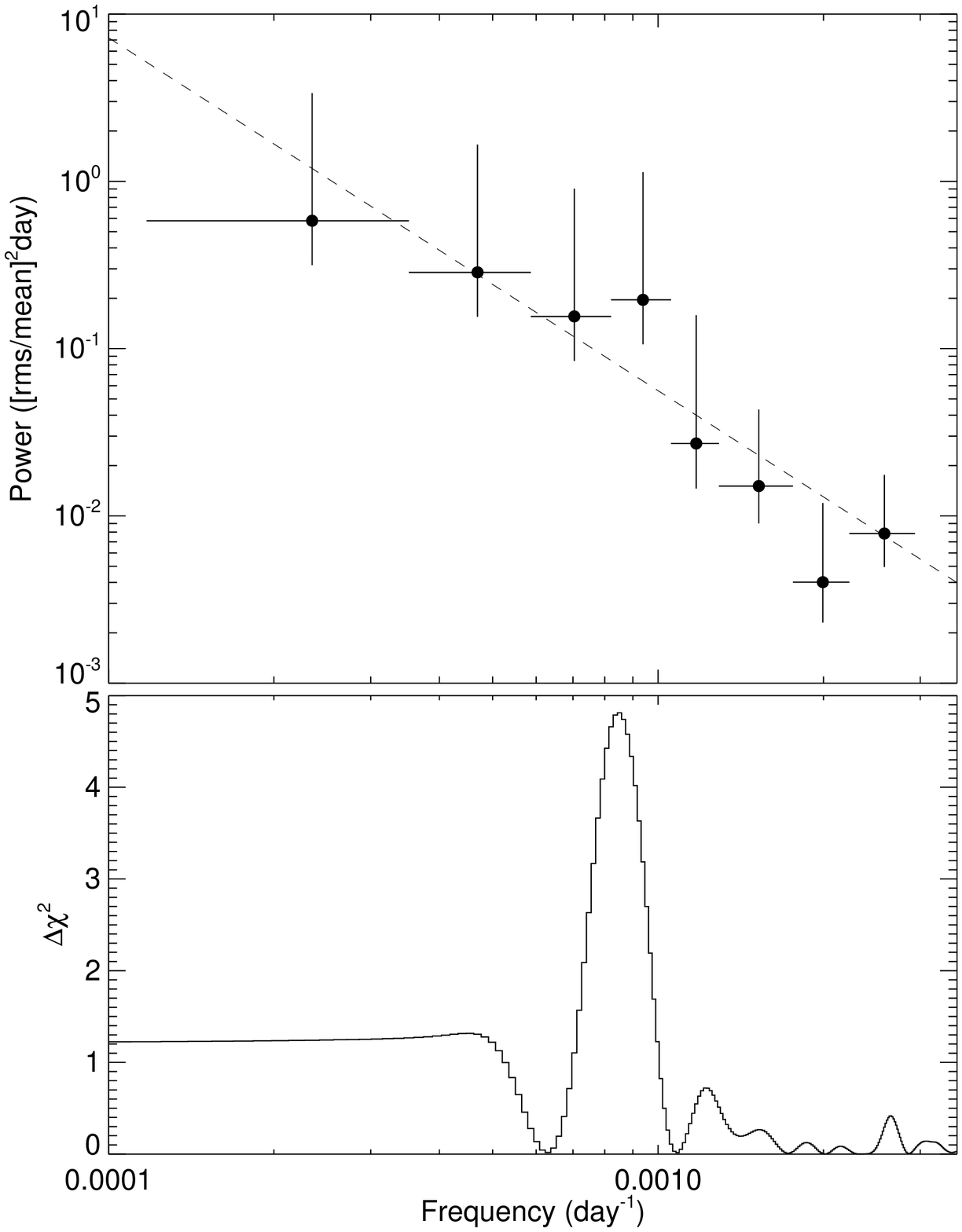}
\caption{(Left): The upper and lower panel show total Crab rates (nebula + pulsar) for layers 2+3 of PCU 2 (black), PCU 3 (red), and PCU 4 (blue) in the 2-15 and 15-50 keV bands, respectively. 
(Right): The upper panel shows the power spectrum of the \rxte\ 15-50 keV rates. The error bars give 68\% confidence intervals. The dashed line is the best-fit power-law.
The lower panel shows the test statistic for a search for periodic signals. \label{rxte}}
\end{figure}

The PCA response has two small time-dependent effects, both accounted for in the response matrices. First, Xenon is slowly accumulating in the front veto layer (nominally filled with Propane) and reducing the low energy sensitivity with time. Second, there is a 
small energy drift in the pulse height channel boundaries, so that a constant channel selection samples a slowly varying energy band. Both effects can influence the rate, though flux determinations (i.e. conversion of count rate to flux) account for 
this. In particular, the correction for changing opacity of the front veto layer is negligible in the 15-50 keV band. Our observed changes in the Crab rate (see Figure~\ref{rxte}) are more than 5 times larger than these effects combined. 


Figure~\ref{rxte} shows total Crab rates for individual \rxte\ pointed observations. From MJD 54690-55435 the Crab rate in PCU 2 declined by $5.1\pm 0.2$\% and $6.8\pm0.3$\% in the 2-15 and 15-50 keV bands, respectively,  relative to MJD 54690.  Similar results, variations of 2-7\%, are seen if the bands are further subdivided. In spectral fits to individual PCA observations, the power law index softens and the normalization and absorption column gradually increase with time, with no clear correlation with flux. These light curves were produced using \rxte/PCA standard 2 data (129 energy channel, 16-second) that were extracted, background subtracted, deadtime corrected using standard \rxte\
recipes\footnote{\url{http://heasarc.gsfc.nasa.gov/docs/xte/recipes/}} and
corrected for the known time dependence of the response.

From visual inspection of the \rxte\ light-curve, three evident peaks
suggest a periodic or quasi-periodic variation with a period of 1000-1500 days. To quantify
these impressions we constructed a power spectrum, shown in the upper right panel of
Figure~\ref{rxte}, and conducted a search for periodic signals. We averaged the corrected 
15-50 keV PCU 2 rates within uniformly spaced bins, using three bins per year, with the yearly interval
where Crab cannot be observed because of Sun constraints occurring in the center of
every third bin. A linear trend, which passed through the first and last binned rate, was
subtracted from the rates, to limit the bleeding of low frequency power into
higher frequency bands.  The power spectrum was then created from the Fourier transform of 
the binned rates.  The lower five points in the plot are from individual Fourier amplitudes,
with the remainder rebinned to reduce errors. A maximum likelihood fit to the unbinned power
spectrum was made using a power-law model. The best fit model is shown, which has a power-law
index of $2.1 \pm 0.4$. 

Standard pulse search methods such as the Lomb test are inappropriate because of the underlying
red noise power spectrum. The test statistic we adopted is the improvement in $\chi^2$ between
fitting the binned rates to a quadratic and to quadratic plus a sinusoid. The quadratic accounts for the low frequency trend in the rates. Since
the source power dominates the counting statistics, we use uniform
errors in the fits, setting $\sigma^2 = P/\Delta t$ where $P$ is the power spectrum model
at the middle of the region where a periodicity may be present ($8.5 \times  10^{-4}$ day$^{-1}$),
and $\Delta t$ the bin width. As seen in the lower panel of Figure~\ref{rxte}, a peak in the
$\Delta \chi^2$ is seen at $(8.5\pm 0.7) \times 10^{-4}$ day$^{-1}$, corresponding to a period of $1180 \pm 100$ days.   However, its significance is only $2\sigma$. A longer history of the Crab flux will be needed to determine if this feature is a property
of the source, or only a statistical fluctuation. Interestingly, this peak value is consistent with twice the period of 568 $\pm$ 10 days found in Crab radio timing noise from 1982 to 1989 \citep{46}.

\begin{figure}
\includegraphics[angle=270]{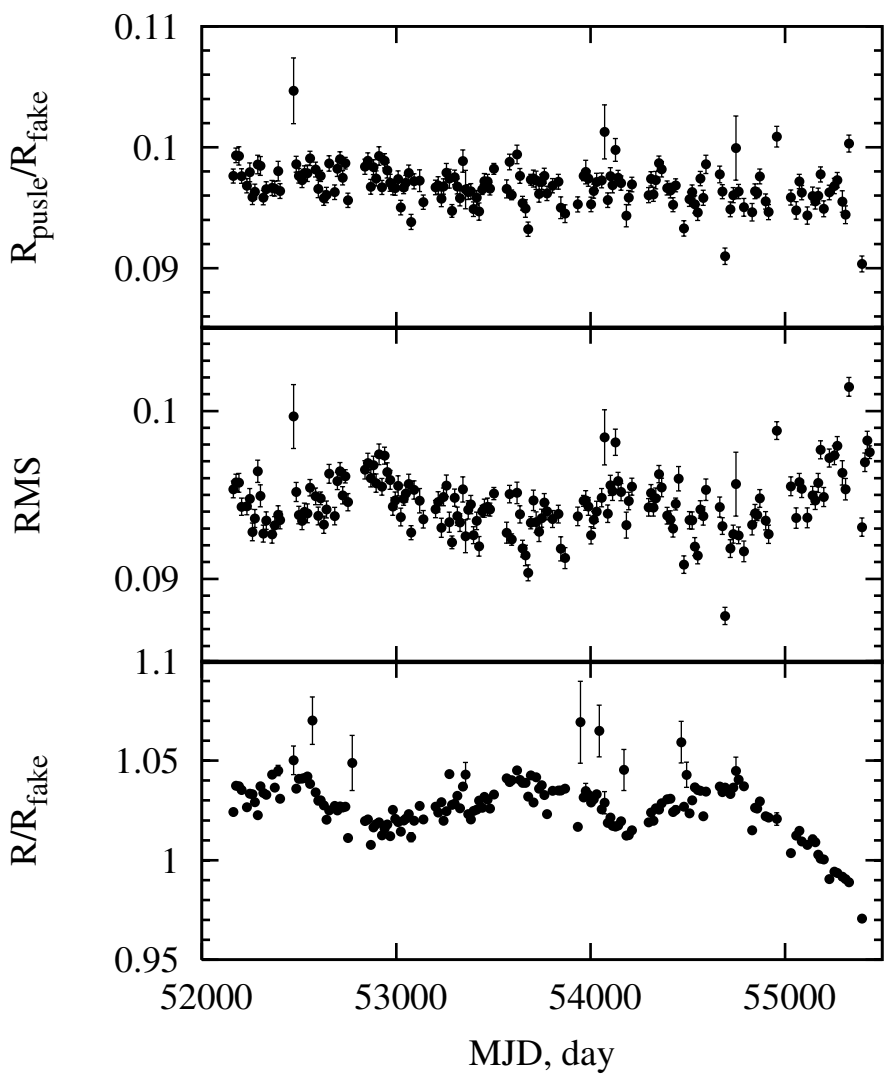}
\caption{(Top): \rxte/PCA pulsed flux (3.2-35 keV). (Center): Fractional Root-mean-squared (RMS) amplitude for the first two harmonics of the pulse period. (Bottom): Total Crab rate in PCU 2. Rates in the top and bottom panels are normalized by the response predicted count rate $R_{\rm fake}$ in the 3.2-35 keV band.\label{rxte_pulsed}}
\end{figure}

The Crab pulsed flux measured using PCU 2 event mode data (250 $\mu$s, 129 energy channels, top layer) is shown in Figure~\ref{rxte_pulsed}. 
Although the pulsed flux (upper panel) steadily decreases at $\sim 0.2$\% yr$^{-1}$, consistent with the pulsar spin-down, the larger (several \% per year) variation in the signal is not seen in the pulsed emission and clearly seems to be nebular in 
origin.

\subsection{\integralsc\ IBIS and JEM-X}

\begin{figure}
\plotone{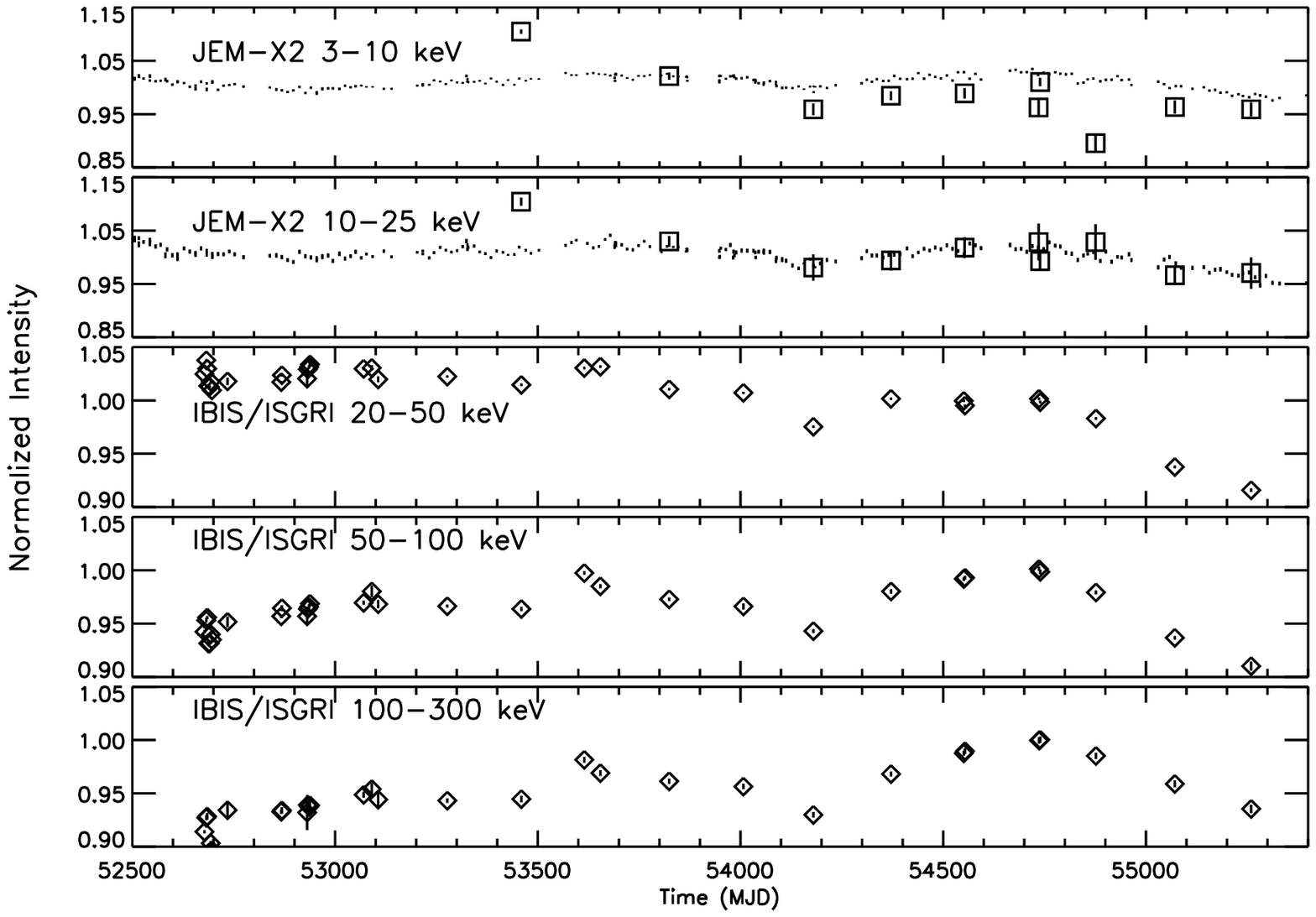}
\caption{\integralsc\ 3-day averaged light curves of the Crab measured in the 3-10 and 10-25 keV bands with JEM-X, and the 20-50, 50-100, and 100-300 keV bands with ISGRI. Normalized \rxte\ PCU2 rates in the 2-15 and 15-50 keV bands are overplotted with the JEM-X data for comparison.\label{integral}}
\end{figure}


Here we present results from the Joint European X-ray Monitor \citep[JEM-X, 3 - 35 keV, ][]{37} and the \integralsc\ Soft Gamma Ray Imager layer of IBIS \citep[ISGRI, 15 keV - 10 MeV, ][]{36} on-board \integralsc\ \citep{34}. The Crab has been observed every spring and fall with \integralsc\ since 2002, mainly for calibration purposes. To reduce systematic effects as much as possible, we have selected on-axis ($<0.25$\arcdeg) observations for JEM-X2 and observations within 10\arcdeg\ of on-axis for IBIS/ISGRI. We include only JEM-X data using the latest on-board software (since MJD 53068).

In Figure~\ref{integral} ISGRI and JEM-X2 count rates from individual pointings averaged over the 3-day \integralsc\ orbit with rms errors are shown. The ISGRI data were analyzed with the Off-line Analysis (OSA) package \citep{40} version 9 with the settings used for the \integralsc\ Galactic Bulge monitoring program\footnote{\url{http://integral.esac.esa.int/BULGE/}} light curves \citep{38}.
Using images from individual pointings, the point-spread function of ISGRI is fitted. These images are integrated in a given energy band after gain, offset, and charge loss corrections are performed for each event. A time-dependent effective area correction, usually performed by assuming that the Crab flux is constant, has been excluded from these data, meaning that not all systematic effects are taken into account. Known effects include residuals in gain and charge loss corrections, present with an amplitude of $\sim1-2$\%, varying on month-years timescales. Similarly, for JEM-X, the ad-hoc piecewise linear correction (added to OSA to reduce time trends in the Crab flux) was excluded from the standard OSA analysis. JEM-X consists of two identical units, JEM-X1 and JEM-X2. During the period of interest, JEM-X2 has mostly been in standby-mode and JEM-X1 the active unit. A gradual decrease has been observed in the sensitivity of JEM-X1, so only JEM-X2 is shown. The scatter in the JEM-X data is large compared to the observed Crab variations, especially below 10 keV. From MJD 54690-55390, the ISGRI 15-50, 50-100, and 100-300 keV flux decreases by $8.2\pm1.1$, $8.3\pm1.1$, and $5.7\pm1.0$\%, respectively, relative to MJD 54690.



\subsection{\swift\ BAT}
\swift/BAT is a coded aperture telescope operating in the 14 - 150 keV range \citep{30}. 
The Swift/BAT 14-50 and 50-100 keV light curves (see Figure~\ref{all}) are based on publicly available 58-month light curves\footnote{\url{http://swift.gsfc.nasa.gov/docs/swift/results/}} from the Swift/BAT all-sky hard X-ray survey \citep{31, Baumgartner10} extended to May 30, 2010 by the BAT team.  
We binned data from individual Swift pointings in 50-day intervals, eliminating pointings of less than 200 seconds duration and those in which less than 15\% of the BAT detectors were illuminated by the Crab.  The statistical errors on each data point  are small (0.1\%) and are 
dominated by systematic errors.  We estimate the systematic errors to be $\sim 0.75$\% by assuming that the long term variations in the lightcurve are due to real variations in the Crab, and that the shorter term variations around that trend are representative of the systematic error. The BAT data show variations in the Crab flux at the level of $\sim 3$\% yr$^{-1}$. From MJD 54690-55340, BAT observes a decrease of $6.7\pm0.7$ and $10.4\pm0.8$\%, in the 15-50 and 50-100 keV bands, respectively, relative to the rate on MJD 54690, similar to the decrease seen by GBM in the same energy range.


\section{Discussion and Summary}

\begin{figure}
\plotone{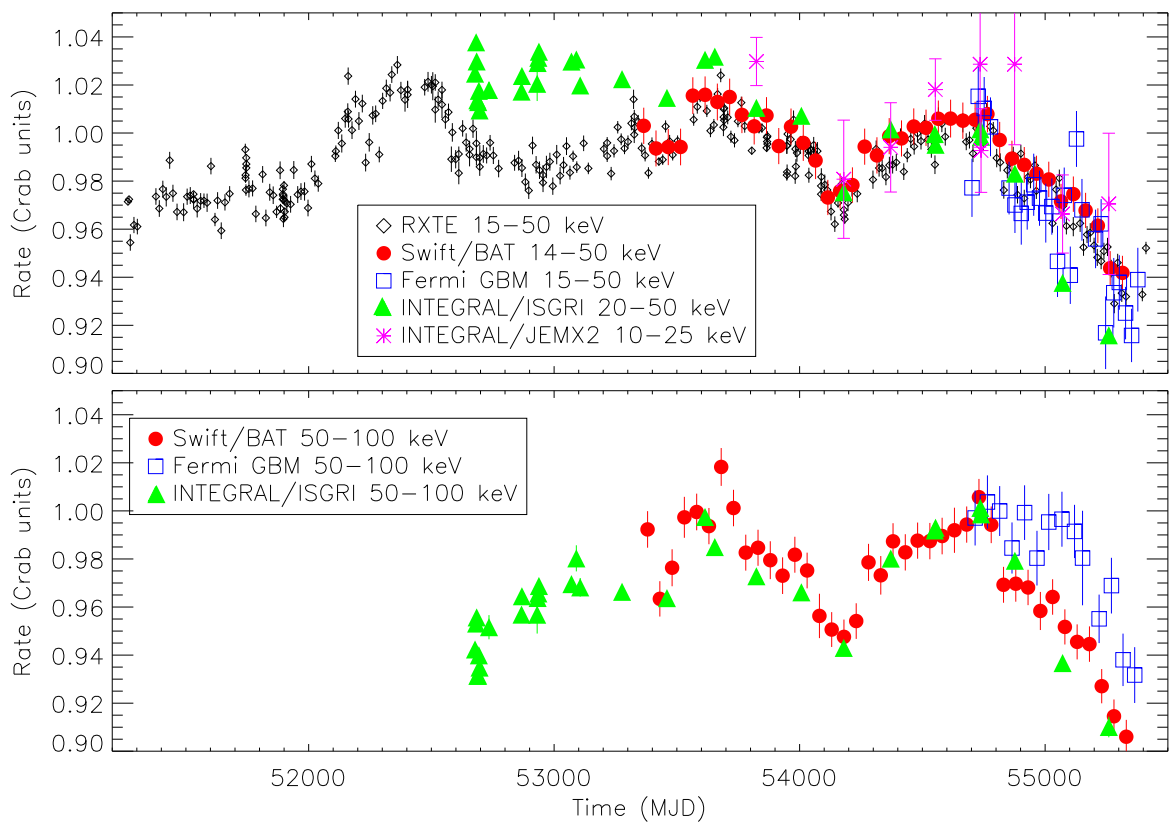}
\caption{Composite Crab light curves for \rxte/PCA (15-50 keV - black diamonds), \swift/BAT (Top: 14-50 keV, Bottom: 50-100 keV - red filled circles), \fermi/GBM (Top: 15-50 keV, Bottom: 50-100 keV - open blue squares), 
 \integralsc/ISGRI (Top: 20-50 keV, Bottom: 50-100 keV - green triangles), and \integralsc/JEM-X2 (10-25 keV). Each data set has been normalized to its mean rate in the time interval MJD 54690-54790. All error bars include only statistical errors.\label{all}} 
\end{figure}

Figure~\ref{all} shows composite light curves combining the overlapping results from \rxte, \integralsc, \swift, and \fermi/GBM. All instruments agree well from
2008 to 2010, with all instruments registering a decline in the Crab 15-50 keV flux of $\sim$7\% (70 mcrab) over the two years starting at MJD 54690, with a similar decline in the 50-100 keV band. PCA and BAT continue to agree back to the start of the \swift\ mission. For \rxte, \swift, and \integralsc/ISGRI the latest measurements shown are significantly below previous minimum. \integralsc/ISGRI shows evidence for the dip near MJD 54100-54200 and the increase before $\sim$MJD 53700, with similar but less significant variations seen in JEM-X2.  Prior to this time, the PCA measurements show continued variations extending back to $\sim$MJD 52000, which are not seen with ISGRI in the 20-50 keV band. We investigated the effect of a change in the default dithering pattern since 2006 March, but found that this cannot explain the observed difference. Known systematic errors in ISGRI energy reconstruction are expected to account for $\sim 1-2$\% deviations. Beginning at $\sim$MJD 54000, there is a strong correlation among the results from the four independent
instruments with very different signal to noise characteristics and observing techniques: Earth occultation, coded-mask imaging, and collimated detectors.  The range of techniques strengthens the case that the
variation is intrinsic to the Crab. We found no apparent correlations between these variations and variations in the \integralsc/SPI anticoincidence detector count rates or GBM count rates, disfavoring local background condition changes as a possible origin, and further supporting a Crab origin. The pulsed flux stability suggests that the observed variations are nebular. 

The observed time variability may be explained by models of the Crab pulsar wind flow.
In some models \citep[e.g.,][]{camus09}, a radial plasma flow in the equatorial plane decelerates 
downstream of a termination shock located at a radius of about 0.5~lyr and near the
inner ring observed in X-rays \citep{5}. Due to adiabatic and synchrotron losses in the fluid the flow becomes inhomogeneous with large variations in local magnetic field strength. These magnetosonic waves are relativistic and the variability timescale is roughly the fluid crossing time across the shock diameter or $1-2$ years. Alternatively \citep{spitkovsky04}, variability on scales of the ion Larmor radius may result from cycles of compression of the electron-positron plasma induced by magnetosonic waves caused by the cyclotron instability in the ion orbits. 
 
{\it Chandra} \citep{5, 52} and {\it XMM-Newton} \citep{53} observations of the Crab
suffer from pile-up effects, making it difficult to monitor absolute fluxes at
the level of a few \%. No {\it Chandra} ACIS observations of the Crab were performed from MJD 54135-55466. Nevertheless, both instruments have shown that the spectrum of the synchrotron X-rays grows distinctly softer as distance from the pulsar increases. Since higher energy electrons have shorter synchrotron lifetimes, the spectrum becomes softer as the particles 
move outward and synchrotron losses grow. Alternatively, the site of the main particle acceleration or the  spectral steepening as a function of distance from the pulsar and shock region may vary with time.	

The differential photon spectrum $dN/dE$ produced by synchrotron-emitting electrons depends on magnetic field strength $B$ and photon energy $E$ as $dN/dE \sim B^{ \gamma } E^{-\gamma }$, where $\gamma$ is the power law photon energy index \citep{54}, suggesting that the observed change in flux could be produced either by a change in the accelerated electron population or a change in the nebular magnetic field of a few percent. 


In summary, the widely-held assumption that the Crab can be used as a standard candle, suitable for normalizing instrument response functions and for calibrating X-ray instruments, should be treated with caution. Although obtaining absolute calibrations and instrument normalizations at $\sim 1$\% is difficult, the results presented here from four independent spacecraft demonstrate that in fact the nebular X-ray/gamma ray emission from the Crab varies at a level of $\sim 3.5$\% yr$^{-1}$. The variation is seen in the nebular emission, and so apparently results from changes in the shock acceleration or the nebular magnetic field. We cannot predict if the present decline will continue or if the $\sim 3$ year pattern will persist. Longer baselines and multi-wavelength observations are needed to answer these questions.

\acknowledgments

 This work is supported by the NASA Fermi 
Guest Investigator program, NASA/Louisiana Board of Regents Cooperative Agreement NNX07AT62A (LSU), the Louisiana Board of Regents Graduate 
Fellowship Program (J. Rodi), and the Spanish Ministerio de Ciencia e Innovaci\'on through the 2008 postdoctoral program MICINN/Fulbright under grant 2008-0116 (A. Camero-Arranz). This research has made use of data obtained through the High Energy Astrophysics Science Archive Research Center Online Service, provided by the NASA/Goddard Space Flight Center; public \swift/BAT  results made available by the \swift/BAT team; and observations with INTEGRAL, an ESA project funded by ESA member states (especially the PI countries: Denmark, France, Germany, Italy, Switzerland, Spain), Poland and with the participation of Russia and the USA.

\clearpage


\begin{thebibliography}{}
\bibitem[Abdo et al.(2010a)]{Abdo10a}
Abdo, A.A. et al. 2010a, \apj, 708, 1254
\bibitem[Aharonian et al.(2004)]{48}
Aharonian, F. et al. 2004, \apj,614, 897
\bibitem[Aharonian et al.(2006)]{55}
Aharonian, F.  et al. 2006, \aap, 457, 899
\bibitem[Albert et al.(2008)]{49}
Albert, J. et al. 2008, \apj, 674, 1037
\bibitem[Aller \& Reynolds (1985)]{19}
Aller, H.D.\& Reynolds, S.P. 1985, \apj, 293, L73
\bibitem[Barthelmy et al.(2005)]{30}
Barthelmy, S.D. et al. 2005, \ssr, 120, 143 
\bibitem[Baumgartner et al.(2010)]{Baumgartner10}
Baumgartner, W. et al. 2010, \apjs, submitted
\bibitem[Bientenholz, Frail, \& Hester (2001)]{13}
Bietenholz, M.F., Frail,  D.A., Hester, J.J. 2001, \apj, 560, 254 
\bibitem[Bissaldi et al.(2009)]{56}
Bissaldi, E. et al.2009, Experimental Astronomy, 24, 47
\bibitem[Camus et al. (2009)]{camus09}
Camus, N.F., Komissarov, S.S., Bucciantini, N., Hughes, P.A., \mnras, 400, 1241
\bibitem[Case et al.(2010)]{28}
Case, G.L. et al. 2010, \apj, submitted, arXiv:1009.4953
\bibitem[Courvoisier et al.(2003)]{40}
Courvoisier, T.J-L. et al. 2003, \aap, 411, L53 
\bibitem[De Jager et al.(1996)]{7}
De Jager, O.C. et al. 1996, \apj, 457 253
\bibitem[Felten \& Morrison (1966)]{54}
Felten, J.E. \& Morrison, P.  1966, \apj, 146, 686
\bibitem[Greiveldinger \& Aschenbach (1999)]{17}
Greiveldinger, C. \& Aschenbach, B. 1999, \apj, 510, 305 
\bibitem[Harmon et al.(2002)]{25}
Harmon, B.A. et al. 2002, \apjs, 138, 149
\bibitem[Hester et al.(1995)]{14}
Hester, J.J. et al. 1995, \apj, 448, 240
\bibitem[Hester et al.(2002)]{15}
Hester, J.J. et al. 2002, \apj, 577, L49 
\bibitem[Hester (2008)]{4}
Hester, J.J. 2008, \araa, 46, 127 
\bibitem[Hoover et al.(2007)]{29}
Hoover, A.S. et al. 2008, in Gamma-Ray Bursts 2007 (AIP Conf. Proc. 1000), eds.  M. Galassi, D. Palmer, E. Fenimore (Melville, NY:AIP), 565 
\bibitem[Jahoda et al.(2006)]{43}
Jahoda, K. et al. 2006, \apjs, 163, 401
\bibitem[Jourdain \& Roques (2009)]{2}
Jourdain, E. \& Roques, J.P. 2009, \apj, 704, 17
\bibitem[Kirsch et al.(2005)]{1}
Kirsch, M.G.F. et al. 2005, \procspie, 5898, 22
\bibitem[Kirsch et al.(2006)]{53}
Kirsch, M.G.F. et al. 2006, \aap, 453, 173
\bibitem[Kuulkers et al.(2007)]{38}
Kuulkers, E. et al. 2007, \aap, 466, 595
\bibitem[Ling \& Wheaton (2003)]{10}
Ling, J.C. \& Wheaton, W.A. 2003, \apj, 598, 334
\bibitem[Ling et al.(2000)]{26}
Ling, J.C. et al. 2000, \apjs, 127, 79
\bibitem[Lund et al.(2003)]{37}
Lund, N. et al. 2003, \aap, 411, L231
\bibitem[Meegan et al.(2009)]{24}
Meegan, C. et al. 2009, \apj, 702, 791
\bibitem[Meyer, Horns, \& Zechlin (2010)]{50}
Meyer, M., Horns,  D., Zechlin, H.-S. 2010, \aap, accepted, arXiV:1008.4524 
\bibitem[Mori et al.(2004)]{52}
Mori, K. et al. 2004, \apj, 609, 186 
\bibitem[Mori et al.(2006)]{16}
Mori, K. et al. 2006, 36th COSPAR Sci. Assembly, Beijing, paper 2615
\bibitem[Much et al.(1995)]{23}
Much, R.  et al. 1995, \aap, 299, 435
\bibitem[Ng \& Romani (2006)]{6}
Ng, C.-Y. \& Romani, R.W. 2006, \apj, 644, 445
\bibitem[Reynolds \& Chevalier (1984)]{20}
Reynolds, S.P. \& Chevalier, R.A. 1984, \apj, 278, 630
\bibitem[Rots, Jahoda, \& Lyne (2004)]{42}
Rots, A., Jahoda,  K., Lyne, A.G. 2004, \apjl, 605, 129
\bibitem[Scott, Finger, \& Wilson (2003)]{46}
Scott, D.M., Finger, M.H., Wilson, C.A. 2003, \mnras, 344, 412
\bibitem[Shaposhnikov (2010)]{44}
Shaposhnikov, N. 2010,  presentation at the 2010 meeting of the IACHEC, Woods Hole, \url{http://web.mit.edu/iachec/meetings/2010/index.html} 
\bibitem[Smith (2003)]{21}
Smith, M. 2003, \mnras, 346, 885
\bibitem[Spitkovsky \& Arons (2004)]{spitkovsky04}
Spitkovsky, A. \& Arons, J. 2004, \apj, 603, 669
\bibitem[Tavani et al.(2010)]{57}
Tavani, M. et al.2010, ATEL \# 2855 
\bibitem[Toor \& Seward (1974)]{12}
Toor, A. \& Seward, F.D. 1974, \aj, 79, 995
\bibitem[Tueller et al.(2010)]{31}
Tueller, J. et al. 2010, \apjs, 186, 378
\bibitem[Ubertini et al.(2003)]{36}
Ubertini, P. et al. 2003, \aap, 411, L131
\bibitem[Verrecchia et al.(2007)]{22}
Verrecchia, F. et al. 2007, \aap, 472, 705
\bibitem[Wakely (2010)]{51}
Wakely, S.P. 2010, Proc. VERITAS Workshop on High Energy Galactic Physics, New York 
\bibitem[Weisskopf et al.(2010)]{3}
Weisskopf, M.C. et al. 2010, \apj, 713, 912 
\bibitem[Weisskopf et al.(2000)]{5}
Weisskopf, M.C. et al.2000, \apj, 536, L81
\bibitem[Wilson-Hodge et al.(2009)]{27}
Wilson-Hodge, C.A. et al. 2009, Proc. Fermi Symposium, eConf C091122, arXiv:0912.3831 
\bibitem[Winkler et al.(2003)]{34}
Winkler, C. et al. 2003, \aap, 411, L1
\end{thebibliography}
\end{document}